\documentclass[aps,prb,reprint,10pt,showpacs,superscriptaddress,floatfix]{revtex4-2}
\usepackage[utf8]{inputenc}
\usepackage[T1]{fontenc}
\usepackage{amsmath,amssymb,physics,bm}
\usepackage{graphicx,booktabs}
\usepackage{orcidlink}
\usepackage{hyperref}
\usepackage[english]{babel}

\newcommand{\wt}{\omega_\tau}
\newcommand{\wc}{\omega_c}

\newcommand{\Tcal}{\mathcal{T}}
\newcommand{\edge}{\mathrm{edge}}
\newcommand{\FP}{\mathrm{FP}}
\newcommand{\EF}{E_F}
\newcommand{\kF}{k_F}
\newcommand{\DeltaZ}{\Delta_Z}
\newcommand{\Geff}{\Gamma_{\mathrm{eff}}}
\newcommand{\Emin}{E^{\mathrm{min}}}

\graphicspath{{figs/}}

\hypersetup{colorlinks=true,linkcolor=blue,citecolor=blue,filecolor=blue,urlcolor=blue,breaklinks=true}

\begin{document}

\title{Torsion-controlled spin transport and tunable intersubband absorption in a screw-dislocated semiconductor nanowire}

\author{Carlos Magno O. Pereira\orcidlink{0000-0002-5170-7538}}
\email[Carlos Magno O. Pereira - ]{carlos.mop@discente.ufma.br}
\affiliation{Programa de P\'os-Gradua\c c\~ao em F\'{\i}sica \& Coordena\c c\~ao do Curso de F\'{\i}sica -- Bacharelado, Universidade Federal do Maranh\~{a}o, 65085-580 S\~{a}o Lu\'{\i}s, Maranh\~{a}o, Brazil}

\author{Edilberto O. Silva\orcidlink{0000-0002-0297-5747}}
\email[Edilberto O. Silva - ]{edilberto.silva@ufma.br}
\affiliation{Programa de P\'os-Gradua\c c\~ao em F\'{\i}sica \& Coordena\c c\~ao do Curso de F\'{\i}sica -- Bacharelado, Universidade Federal do Maranh\~{a}o, 65085-580 S\~{a}o Lu\'{\i}s, Maranh\~{a}o, Brazil}

\date{\today}

\begin{abstract}
We develop an effective-mass theory for spin-resolved transport and intersubband optical absorption in a finite semiconductor nanowire containing a uniform screw-dislocation-induced torsion. The central transport quantity is the band bottom of each torsion-dependent one-dimensional subband: torsion lowers this band bottom while an axial magnetic field resolves the two Zeeman branches, so that a spin-selective torsion interval opens when the Fermi level lies between the two branch minima. We distinguish this band-bottom transport threshold from the fixed-momentum transverse energy that governs the vertical intersubband optical transition, and obtain a design rule for the width of the spin-selective torsion interval that is linear in the Zeeman splitting in the small-splitting limit. Finite-size effects are incorporated by treating the active region as an open scattering region of length $L$ and radius $R$, with Fabry--Perot interference, radial boundary conditions, and surface-induced linewidth broadening. The spin-resolved response remains well defined when the effective linewidth satisfies $\Gamma_{\mathrm{eff}}\ll\Delta_Z$, a criterion quantified through polarization and conductance-contrast maps. In the optical sector, finite-radius intersubband absorption gives a torsion-tunable resonance in the THz range; surface boundary conditions shift the absolute resonance frequency and modify the oscillator strength, while, for the dominant unit-branch transition, the leading fixed-momentum torsional slope $\partial\omega/\partial\tau\simeq\hbar k_F/m^*$ is preserved. The resulting framework connects geometric torsion, spin-resolved mesoscopic transport, spin thermopower, and tunable intersubband absorption within a single open-quantum-wire model.
\end{abstract}

\maketitle

\section{Introduction}
\label{sec:intro}

The coupling between geometry and quantum dynamics provides a fertile ground for spectral and transport engineering. In low-dimensional systems, strain, curvature, and topological defects can be encoded as effective gauge fields or geometry-induced potentials acting on the charge carriers~\cite{Jensen1971,daCosta1981,Ortix2015}. Line defects in crystalline media admit a geometric description in terms of torsion and curvature fields, as established in the Riemann--Cartan formulation of defects in solids~\cite{Katanaev1992,Vozmediano2010}. Screw dislocations, in particular, generate an effective torsional geometry along the growth axis of a nanowire and can be described within geometric elasticity, where the metric encodes the lattice distortion and the corresponding confinement experienced by the electron gas.

In parallel, the discovery of moir\'e superlattices in twisted van der Waals heterostructures has led to the field of \emph{twistronics}, in which the twist angle between stacked layers is used as a control knob for band engineering and correlated phases. Continuum descriptions of rotated graphene layers established twist as an electronic band-structure parameter~\cite{LopesDosSantos2007,Mele2010}, while the magic-angle theory showed that small-angle stacking can generate flat moir\'e bands~\cite{Bistritzer2011}. Experiments on magic-angle graphene superlattices subsequently revealed correlated insulating states and unconventional superconductivity~\cite{Cao2018Insulator,Cao2018}. These results established mechanical rotation as a powerful band-structure control parameter in two-dimensional materials and motivated analogous geometric control strategies in lower-dimensional conductors.

Spintronics aims at exploiting the spin degree of freedom of carriers for non-volatile memories, logic, and quantum information processing~\cite{Wolf2001,Zutic2004,Awschalom2007}. Traditional architectures rely on ferromagnetic contacts, spin-dependent scattering, and spin-orbit coupling to generate and detect spin-polarized currents. More recently, spin caloritronics has highlighted the coupling between spin and heat transport, including thermally driven spin torques, spin Seebeck signals, and thermally generated spin injection~\cite{Hatami2007,Uchida2008,Xiao2010,Slachter2010,Bauer2012}. However, in most of these platforms, the relevant control parameters are magnetic fields, gate voltages, or temperature, while purely mechanical control of spin and spin caloritronic responses remains comparatively unexplored.

From the geometric side, several works have shown that curvature, strain, and torsion in deformed nanostructures can strongly modify subband structure and transport, effectively acting as built-in confinement, synthetic gauge fields, or pseudomagnetic fields for orbital motion~\cite{Guinea2010,SilvaNetto2008,Cuoghi2011}. Screw-dislocation growth and Eshelby twist in nanowires provide a concrete materials context in which torsional deformation and surface energetics are intertwined~\cite{Akatyeva2012}, so that the torsion densities considered here are accessible in wires grown with an embedded screw dislocation or subjected to a controlled mechanical twist. Within the geometric elasticity approach, screw dislocations are modeled by a Riemann--Cartan manifold where torsion couples to the electronic wave function and generates a torsion-induced radial confinement in screw-dislocated nanowires. Closely related torsion-based models have recently been used to describe geometry-induced chiral currents in helicoidal quantum wells, screw-dislocation-engineered quantum dots with torsion-tunable nonlinear optical response, nonlinear optical resonances of confined electrons under torsion and magnetic fields, and geometrical optical activity in media with a continuous distribution of screw dislocations~\cite{Silva2026HelicoidalQW,Silva2026ScrewQuantumDot,PereiraSilva2026NonlinearOpticalTorsion,BelichSilva2026OpticalActivity}. In the spin-degenerate limit, this mechanism can be viewed as a geometric gate: a uniform torsion density displaces the orbital subband bottoms and thereby modifies conductance steps, thermoelectric coefficients, and shot-noise signatures within the Landauer framework. Crucially, torsion is a \emph{non-electrostatic} control knob: unlike a gate voltage, it tunes the subband bottoms without injecting charge, without the associated capacitive loading, and through a geometric coupling to the same orbital index $A_{nm}$ that an electrostatic field cannot reproduce. It is therefore complementary to gating, operative where electrostatic depletion is impractical, and naturally compatible with mechanical or nanoelectromechanical actuation. This establishes torsion-controlled transport as a geometric analog of quantum point contacts controlled by twist rather than by electrostatic depletion.

Here we extend this framework to spin-resolved transport and optical response. By incorporating the Pauli interaction with an axial magnetic field, we demonstrate that the interplay between torsion-induced orbital shifts and Zeeman splitting lifts the spin degeneracy and produces a torsion-tunable spin-resolved threshold structure in a screw-dislocated nanowire. The central new ingredient relative to the spin-degenerate case is that torsion and the Zeeman term act on distinct quantum numbers; torsion lowers the orbital band bottoms while the field splits the two spin branches, so that a purely mechanical twist opens a \emph{spin-selective} window of channel opening that has no analog in the degenerate limit. At the same time, torsion renormalizes the transverse energy levels and, hence, the intersubband spacing, thereby shifting the absorption resonance across the THz regime. Building on the Landauer--B\"uttiker formalism and the standard theory of mesoscopic thermoelectric transport~\cite{Landauer1957,Buttiker1986,Sivan1986,Butcher1990,Blanter2000,Beenakker1997,HicksDresselhaus1993b,MahanSofo1996}, we formulate the transport problem as a finite open scattering region rather than a closed quantum dot. This requires a careful distinction between two physically different quantities: the band bottom of the torsion-dependent subband, which sets the Landauer channel-opening threshold, and the fixed-momentum transverse energy at the Fermi wave vector, which sets the vertical intersubband optical transition. The distinction is essential because a Landauer measurement probes the existence of propagating modes, controlled by the band bottom, in a narrow energy window around the reservoir chemical potential, whereas optical absorption is a vertical transition at the occupied longitudinal momentum. We then discuss how finite device length, finite radius, and realistic boundary scattering modify the ideal open-channel picture. The resulting effect is accessible through two complementary experimental channels, spin-resolved conductance steps and a torsion-tunable THz absorption resonance, together with a torsion-reversible spin thermopower, all controlled by the same geometric parameter. The analysis identifies a common threshold mechanism by which geometric torsion controls spin, thermoelectric, and optical responses in one-dimensional nanostructures.

\section{Spin-dependent spectrum}
\label{sec:spin}

\subsection{Zeeman splitting and torsion-dependent dispersion}

The effective-mass Hamiltonian in the presence of an axial magnetic field $B\hat{z}$ and uniform torsion $\tau$, including the spin interaction, is written as
\begin{equation}
    H = H_{\mathrm{orb}} + H_{\mathrm{Zeeman}}
      = H_{\mathrm{orb}} + \frac{1}{2} g^* \mu_B \sigma_z B,
\end{equation}
where $H_{\mathrm{orb}}$ is the orbital Hamiltonian derived in geometric elasticity theory, $g^*$ is the effective Land\'e factor, $\mu_B$ is the Bohr magneton, and $\sigma_z=\pm1$ denotes the spin projection. It is useful to define the orbital index factor
\begin{equation}
    A_{nm}=n+\frac{|m-\ell|}{2}-\frac{m-\ell}{2}+\frac12,
    \label{eq:Anm}
\end{equation}
which labels the torsion-renormalized transverse branch. Here $n=0,1,2,\dots$ is the radial quantum number, $m\in\mathbb{Z}$ is the orbital angular-momentum index, and $\ell$ is the dislocation-induced angular-momentum shift, an effective flux fixed by the Burgers vector of the embedded screw dislocation, which enters $A_{nm}$ only through the integer combination $m-\ell$. It is important to distinguish between the two geometric labels in the model, since they play different physical roles and are treated as independent control parameters. The parameter $\ell$ is the \emph{discrete} topological charge carried by a dislocation line threading the wire; it is a fixed structural property of a given sample and enters solely through $A_{nm}$. The torsional density $\tau$, by contrast, is the \emph{continuous} effective deformation entering the electronic metric through the $g_{z\varphi}$ cross term, and is the quantity varied throughout this work (for instance by an applied Eshelby twist or a mechanically imposed torsion). Sweeping $\tau$ at fixed $A_{nm}$ therefore corresponds physically to mechanically twisting a wire that carries a fixed dislocation content, rather than to changing the dislocation charge itself; the two would coincide only in the special case of a uniformly twisted rod whose twist is generated entirely by a single line defect, which is not assumed here. This convention follows the effective-geometric viewpoint used in earlier helicoidal and screw-dislocation models of torsion-induced confinement and optical response~\cite{Silva2026HelicoidalQW,Silva2026ScrewQuantumDot,PereiraSilva2026NonlinearOpticalTorsion}. The spin-resolved dispersion can then be written as
\begin{align}
    E_{n,m,\sigma}(k_z;\tau)
    &= \hbar(\omega_\tau+\omega_c)A_{nm}
       +\frac{\hbar^2k_z^2}{2m^*}
       +\frac{\sigma}{2}\DeltaZ,
       \label{eq:Enm_spin}\end{align}
where $\omega_\tau=\hbar k_z\tau/m^*$, $\omega_c=eB/m^*$, $\DeltaZ=|g^*|\mu_BB$ is the Zeeman splitting magnitude, and $\sigma=+1$ ($-1$) labels the higher- (lower-) energy Zeeman branch. The sign of $g^*$ fixes which spin projection corresponds to each branch. Equivalently,
\begin{align}
    E_{n,m,\sigma}(k_z;\tau)
    &=\hbar\omega_cA_{nm}+\frac{\sigma}{2}\DeltaZ
      +\frac{\hbar^2}{2m^*}\left(k_z+\tau A_{nm}\right)^2
      \notag\\
    &\quad -\frac{\hbar^2\tau^2A_{nm}^2}{2m^*}.
    \label{eq:completed_square}
\end{align}
Equation~\eqref{eq:completed_square} makes explicit that torsion changes not only the transverse confinement energy at fixed longitudinal momentum but also the location of the minimum of the longitudinal dispersion. We emphasize that the torsional contribution to the transverse energy, $\hbar\omega_\tau A_{nm}=\hbar^2 k_z\tau A_{nm}/m^*$, is itself momentum dependent: it originates from the $g_{z\varphi}$ cross term of the screw-dislocation metric and therefore acts as a dynamical, $k_z$-coupled confinement rather than a static transverse potential. This observation is important for a consistent interpretation of transport.

\subsection{Open-channel thresholds: band bottom versus fixed-momentum energy}
\label{subsec:kz_thresholds}

Two distinct energy scales are associated with Eq.~\eqref{eq:Enm_spin}, and they play different physical roles. The first is the global band minimum of the one-dimensional subband,
\begin{equation}
    \Emin_{n,m,\sigma}(\tau)
    =\hbar\omega_cA_{nm}+\frac{\sigma}{2}\DeltaZ
     -\frac{\hbar^2\tau^2A_{nm}^2}{2m^*},
    \label{eq:true_minimum}
\end{equation}
obtained at $k_z=-\tau A_{nm}$. In a two-terminal Landauer geometry this is the relevant \emph{channel-opening threshold}: a propagating mode at energy $E$ exists only when $E>\Emin_{n,m,\sigma}$, which is precisely the positivity condition $\mathcal{D}_\nu(E,\tau)>0$ of the longitudinal roots in Eq.~\eqref{eq:k_roots} below, so the spin-resolved channel conducts when $\EF>\Emin_{n,m,\sigma}$. Because torsion enters $\Emin$ through $-\hbar^2\tau^2A_{nm}^2/2m^*$, it \emph{lowers} the band bottom; increasing torsion therefore drives channels open rather than closed.

A spin-resolved channel thus opens at the torsion value where $\Emin_{n,m,\sigma}(\tau)=\EF$,
\begin{equation}
    \tau^{\mathrm{open}}_{n,m,\sigma}
    =\frac{1}{A_{nm}}\sqrt{\frac{2m^*}{\hbar^2}
      \left(\hbar\omega_cA_{nm}+\frac{\sigma}{2}\DeltaZ-\EF\right)},
    \label{eq:tau_open}
\end{equation}
which requires the band bottom to lie above $\EF$ at zero torsion, $\hbar\omega_cA_{nm}+\tfrac{\sigma}{2}\DeltaZ>\EF$. The lower-energy branch reaches $\EF$ at the smaller torsion, so the width of the spin-selective torsion interval is
\begin{align}
    \Delta\tau_{n,m}
    &=\frac{\sqrt{2m^*}}{\hbar A_{nm}}
      \left[\sqrt{\hbar\omega_cA_{nm}+\tfrac{\DeltaZ}{2}-\EF}
      \right.\notag\\
    &\qquad\left.-\sqrt{\hbar\omega_cA_{nm}-\tfrac{\DeltaZ}{2}-\EF}\,\right]
      \label{eq:spin_window}\\
    &\xrightarrow[\DeltaZ\ll\,\hbar\omega_cA_{nm}-\EF]{}
      \frac{\sqrt{2m^*}\,\DeltaZ}
           {2\hbar A_{nm}\sqrt{\hbar\omega_cA_{nm}-\EF}}.
      \notag
\end{align}
The interval grows with the Zeeman splitting and, in the small-splitting limit, is linear in $\DeltaZ$ and decreases with $A_{nm}$. This expression summarizes the basic mechanism: torsion lowers the bottom of the orbital channel, while the axial magnetic field splits the two spin thresholds.

The second scale is the transverse channel energy evaluated at the reservoir-selected Fermi momentum,
\begin{equation}
    \varepsilon^{\mathrm{ch}}_{n,m,\sigma}(\tau;\kF)
    \equiv E_{n,m,\sigma}(\kF;\tau)-\frac{\hbar^2\kF^2}{2m^*},
    \label{eq:channel_threshold_def}
\end{equation}
which gives
\begin{align}
    \varepsilon^{\mathrm{ch}}_{n,m,\sigma}(\tau;\kF)
    &=\varepsilon^0_{n,m}+\alpha_{n,m}\tau+\frac{\sigma}{2}\DeltaZ,
    \label{eq:lead_threshold}\\
    \varepsilon^0_{n,m}&=\hbar\omega_cA_{nm},
    \qquad
    \alpha_{n,m}=\frac{\hbar^2\kF}{m^*}A_{nm}.
    \notag
\end{align}
This fixed-$k_z$ quantity is \emph{not} the transport threshold; it is introduced because the intersubband optical transition is a vertical (momentum-conserving) excitation evaluated at the occupied longitudinal momentum $k_z\simeq\kF$ (Sec.~\ref{sec:optics}). We retain $\varepsilon^{\mathrm{ch}}_{n,m,\sigma}$ solely for that optical resonance, where, for the difference between two transverse branches, the $\tau^2$ terms of Eq.~\eqref{eq:completed_square} cancel and only the linear torsional slope survives.

In the presence of radial boundary shifts and linewidth broadening, the transport threshold becomes
\begin{equation}
    \Emin{}^{,\edge}_{n,m,\sigma}
    =\hbar\omega_cA_{nm}+\frac{\sigma}{2}\DeltaZ
     -\frac{\hbar^2\tau^2A_{nm}^2}{2m^*}
     +\delta\varepsilon_{n,m}(R,\lambda_s,\tau),
    \label{eq:lead_threshold_edge}
\end{equation}
where $\delta\varepsilon_{n,m}$ is the finite-radius correction discussed below. If the ideal threshold is broadened by an effective width
\begin{equation}
    \Geff\simeq\sqrt{\Gamma_{\edge}^2+(\eta k_BT)^2},
    \label{eq:Gamma_eff}
\end{equation}
with $\eta$ a number of order unity set by the chosen linewidth convention, the spin contrast near a threshold is controlled by the ratio $\DeltaZ/\Geff$. A useful estimate is
\begin{equation}
    \frac{G_{\mathrm{low}}}{G_{\mathrm{high}}}\sim
    \exp\left(\frac{\DeltaZ}{\Geff}\right),
    \label{eq:contrast_estimate}
\end{equation}
which makes the robustness criterion explicit: a well-defined spin-selective interval requires $\Geff\ll\DeltaZ$. Equation~\eqref{eq:contrast_estimate} is a thermally activated (tail) estimate valid when both spin-resolved thresholds lie above $\EF$, so that each branch conductance is in the exponentially activated regime; near and below the thresholds, the ratio saturates, and Eq.~\eqref{eq:contrast_estimate} provides only an upper bound on the attainable contrast. This criterion will reappear in the finite-boundary analysis and in the interpretation of the transistor-like transfer curves.
\begin{figure}[t]
  \centering
  \includegraphics[width=0.98\linewidth]{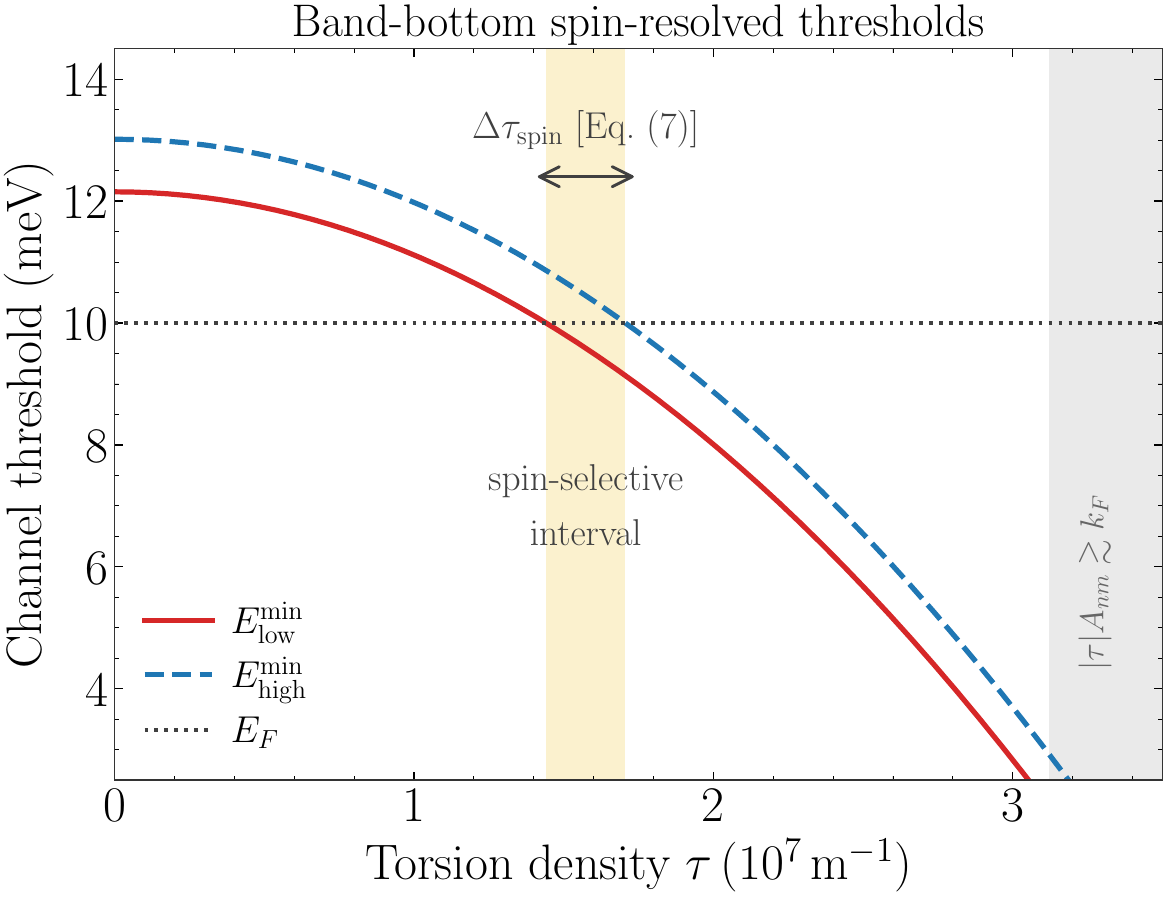}
  \caption{\textbf{Spin-resolved channel-threshold mechanism.}
  Torsion lowers the band bottom $\Emin_\nu(\tau)$ of each subband through the term $-\hbar^2\tau^2A_{nm}^2/2m^*$, while the Zeeman term separates the low- and high-energy branches by $\Delta_Z$. A spin-selective torsion interval appears when the Fermi level lies between the two branch minima. Its width is set by the scale $\Delta\tau_{\mathrm{spin}}$ of Eq.~\eqref{eq:spin_window}, which is linear in $\Delta_Z$ in the small-splitting limit. The labels ``low'' and ``high'' denote the two Zeeman-split energy branches; the association with spin-up or spin-down is fixed by the sign of $g^*$.}
  \label{fig:threshold_mechanism}
\end{figure}

\begin{table}[t]
\caption{Representative parameter set used in the numerical illustrations. The values are chosen to be typical of narrow-gap III--V nanowires with a large effective $g$ factor, using InAs material scales as a reference~\cite{Vurgaftman2001}. The robustness maps are expressed in terms of the dimensionless ratio $\Gamma_{\mathrm{edge}}/\Delta_Z$, so that the robustness conclusions can be rescaled to other materials and magnetic fields.}
\label{tab:parameters}
\begin{ruledtabular}
\begin{tabular}{lll}
Quantity & Symbol & Value or range \\
\hline
Effective mass & $m^*$ & $0.023\,m_e$ \\
Effective $g$ factor & $|g^*|$ & $14.9$ \\
Axial magnetic field & $B$ & $1$--$2~\mathrm{T}$ \\
Temperature & $T$ & $0.5$--$8~\mathrm{K}$ \\
Fermi energy & $E_F$ & $10~\mathrm{meV}$ \\
Torsion density & $\tau$ & $0$--$6\times10^7~\mathrm{m}^{-1}$ \\
Active length & $L$ & $0.1$--$1~\mu\mathrm{m}$ \\
Wire radius & $R$ & $20$--$80~\mathrm{nm}$ \\
Surface linewidth & $\Gamma_{\mathrm{edge}}$ & varied relative to $\Delta_Z$ \\
\end{tabular}
\end{ruledtabular}
\end{table}

\subsection{Model assumptions and regime of validity}
\label{subsec:validity}

The analysis above defines a controlled, effective model rather than a microscopic solution of all electronic, elastic, and electrostatic degrees of freedom of a real nanowire. Its range of validity is set by several independent scale separations. First, the geometric-elastic description of a screw-dislocated medium assumes a continuum limit: the torsional length scale must remain large compared with the lattice spacing $a$, so that
\begin{equation}
    |\tau|a \ll 1.
    \label{eq:continuum_condition}
\end{equation}
For III-V lattice constants, this condition is well satisfied for the torsion densities considered here. A separate, more restrictive criterion applies if $\tau$ is interpreted as a uniform macroscopic twist of the entire cylindrical wire. In that case, the dimensionless surface strain is controlled by
\begin{equation}
    |\tau|R,
    \label{eq:tauR_parameter}
\end{equation}
with a conservative small-strain regime corresponding to $|\tau|R\ll1$ and a qualitative upper bound of order $|\tau|R\lesssim1$. For the representative radii in Table~\ref{tab:parameters}, this implies that the largest values displayed on the torsion axis should be understood as the strong-torsion part of the effective parameter space, especially for larger radii. In a screw-dislocation background, however, $\tau$ denotes the effective torsional density entering the electronic metric rather than necessarily the twist angle per unit length of a uniformly twisted elastic rod. The most conservative interpretation is therefore to regard the high-$\tau$ region as a parametric extrapolation used to expose the threshold mechanism, while the small-strain subset is obtained by restricting $|\tau|R$ below unity.

Second, the band-bottom lowering $-\hbar^2\tau^2A_{nm}^2/2m^*$ that drives the channel-opening mechanism is obtained within the effective-mass parabolic dispersion, which is controlled only when the torsional shift of the longitudinal wave vector is not larger than the incident Fermi scale,
\begin{equation}
    |\tau|A_{nm}\lesssim k_F,
    \label{eq:kF_condition}
\end{equation}
with the strictly perturbative limit given by $|\tau|A_{nm}\ll k_F$. For the parameter set in Table~\ref{tab:parameters}, $E_F=10~\mathrm{meV}$ and $m^*=0.023m_e$ give $k_F\simeq7.8\times10^7~\mathrm{m}^{-1}$. Thus, the plotted torsion interval reaches the regime where torsion-induced orbital shifts become comparable to the Fermi momentum scale for the lowest branches. Beyond this scale, the quadratic lowering of $\Emin$ would formally drive the band bottom to arbitrarily negative values and the open-channel count would grow without bound; that part of the torsion axis is therefore to be read as a parametric extrapolation, and the controlled spin-selective behavior is the one occurring within Eq.~\eqref{eq:kF_condition}.

Third, the spin structure is treated at the Pauli--Zeeman level. Spin-orbit coupling (SOC), including the Rashba and Dresselhaus terms, is not explicitly included. This approximation isolates the torsion--Zeeman threshold mechanism and is controlled when the spin--orbit splitting at the Fermi momentum is small compared with the Zeeman splitting,
\begin{equation}
    E_{\mathrm{SO}}\simeq\alpha k_F \ll \Delta_Z,
    \label{eq:SO_condition}
\end{equation}
and when the relevant orbital branches are well separated. It is important to assess this hierarchy quantitatively for the reference material, since the large effective $g$ factor of narrow-gap III--V nanowires (the very feature that produces a sizeable $\Delta_Z$) is generically accompanied by strong structural-inversion-asymmetry SOC. For an InAs nanowire with a representative Rashba coefficient $\alpha\simeq20~\mathrm{meV\,nm}$~\cite{Fasth2007} and the Fermi momentum $k_F\simeq7.8\times10^7~\mathrm{m}^{-1}$ of Table~\ref{tab:parameters}, one finds $E_{\mathrm{SO}}\simeq\alpha k_F\approx1.6~\mathrm{meV}$, which is comparable to the Zeeman splitting $\Delta_Z=|g^*|\mu_BB\approx1.3$--$1.7~\mathrm{meV}$ over the field range $B=1.5$--$2~\mathrm{T}$. The Pauli--Zeeman description used here is therefore quantitatively controlled in a delimited window, higher axial field, or a gate-reduced/screened Rashba field, or a host with large $g^*$ but weaker SOC, rather than for arbitrary parameters. In the perturbative subregime $\alpha k_F\ll\Delta_Z$, the leading effect of an axial-field/azimuthal Rashba term linear in $k_z$ is not to close the spin-selective interval but to renormalize the band curvature of the two branches,
\begin{equation}
    \frac{\hbar^2}{2m^*}\;\longrightarrow\;
    \frac{\hbar^2}{2m^*_\pm},
    \qquad
    \frac{\hbar^2}{2m^*_\pm}=\frac{\hbar^2}{2m^*}\pm\frac{\alpha^2}{\Delta_Z},
    \label{eq:mass_renorm}
\end{equation}
obtained by expanding $\tfrac{1}{2}\sqrt{\Delta_Z^2+(2\alpha k_z)^2}\simeq\tfrac{1}{2}\Delta_Z+\alpha^2 k_z^2/\Delta_Z$ for the upper branch and with the opposite sign for the lower branch. Here, the upper (lower) sign corresponds to the higher- (lower-) energy Zeeman branch, so that the higher branch is slightly stiffened and the lower branch slightly softened, independently of the sign convention chosen for $g^*$. The separation between the two band bottoms remains $\approx\Delta_Z$ to leading order while the branches acquire slightly different effective masses. Outside this window, $\sigma_z$ and the SOC field no longer commute, spin remains a mixed quantum number, and the present low/high-branch description should be replaced by a full spinor scattering matrix; the qualitative threshold idea should survive, but the branch polarization and the optical selection rules would be modified by spin--orbit-induced hybridization. A detailed treatment of the competition between torsion and SOC, in particular the fact that both enter as terms linear in $k_z$, one coupled to the orbital index $A_{nm}$ and the other to spin, is left for future work.

Finally, the transport model assumes a noninteracting, phase-coherent, or quasi-ballistic conductor in linear response. Electron-electron interactions, Coulomb blockade, self-consistent electrostatics, piezoelectric fields, electron-phonon relaxation, and microscopic disorder configurations are not solved explicitly. Their leading effect at the level of the present open-channel theory is represented by the effective linewidth $\Gamma_{\mathrm{eff}}$ and by the surface-limited mean free path introduced below. Similarly, the optical calculation uses a minimal Lorentzian intersubband absorption model; depolarization shifts, excitonic corrections, and a fully microscopic evaluation of the finite-radius dipole matrix elements would refine the line shape and oscillator strength but not the leading torsional slope. With these assumptions, the operative hierarchy of the model can be summarized as
\begin{equation}
    |\tau|a\ll1,
    \;\;
    |\tau|R\lesssim1,
    \;\;
    |\tau|A_{nm}\lesssim k_F,
    \;\;
    \Gamma_{\mathrm{eff}}\ll\Delta_Z,
    \label{eq:validity_hierarchy}
\end{equation}
with $E_{\mathrm{SO}}\simeq\alpha k_F\ll\Delta_Z$ when the Zeeman branches are to be interpreted as approximately spin resolved.

\subsection{Numerical consistency check of the effective threshold}
\label{subsec:numerical_benchmark}

Before using Eq.~\eqref{eq:true_minimum} as the channel-opening threshold in the transport calculation, it is useful to check explicitly how this band bottom emerges from the torsion-dependent dispersion. This check is not intended to be a microscopic finite-radius simulation of the full nanowire, nor does it replace the surface correction $\delta\varepsilon_{n,m}(R,\lambda_s,\tau)$ introduced above. Rather, it is a numerical consistency test of the effective open-channel spectrum itself, designed to verify that the Landauer threshold is obtained by minimizing the dispersion with respect to $k_z$, instead of evaluating the transverse energy at the fixed reservoir momentum $k_F$.

For a fixed transverse branch $A_{nm}$ and Zeeman branch $\sigma$, we numerically minimize
\begin{equation}
    E^{\mathrm{eff}}_{n,m,\sigma}(k_z;\tau)
    = \hbar\omega_c A_{nm}
    +\frac{\sigma}{2}\Delta_Z
    +\frac{\hbar^2 k_z^2}{2m^*}
    +\frac{\hbar^2 k_z\tau A_{nm}}{m^*},
    \label{eq:numerical_benchmark_dispersion}
\end{equation}
with respect to the longitudinal wave vector. The resulting numerical threshold is
\begin{equation}
    E^{\mathrm{min,num}}_{n,m,\sigma}(\tau)
    = \min_{k_z} E^{\mathrm{eff}}_{n,m,\sigma}(k_z;\tau).
    \label{eq:Emin_num}
\end{equation}
The analytic minimization gives $k_z^{\mathrm{min}}=-\tau A_{nm}$ and therefore
\begin{equation}
    E^{\mathrm{min,an}}_{n,m,\sigma}(\tau)
    = \hbar\omega_c A_{nm}
    +\frac{\sigma}{2}\Delta_Z
    -\frac{\hbar^2\tau^2A_{nm}^2}{2m^*}.
    \label{eq:Emin_an_benchmark}
\end{equation}
Thus, after subtracting the zero-torsion value, the expected band-bottom shift is
\begin{equation}
    \Delta E^{\mathrm{min}}(\tau)
    \equiv E^{\mathrm{min}}(\tau)-E^{\mathrm{min}}(0)
    =-\frac{\hbar^2\tau^2A_{nm}^2}{2m^*}.
    \label{eq:benchmark_shift}
\end{equation}
This is the same quadratic lowering that appears in Eq.~\eqref{eq:true_minimum} and is responsible for the torsion-driven opening of transport channels.

\begin{figure}[t]
  \centering
  \includegraphics[width=0.98\linewidth]{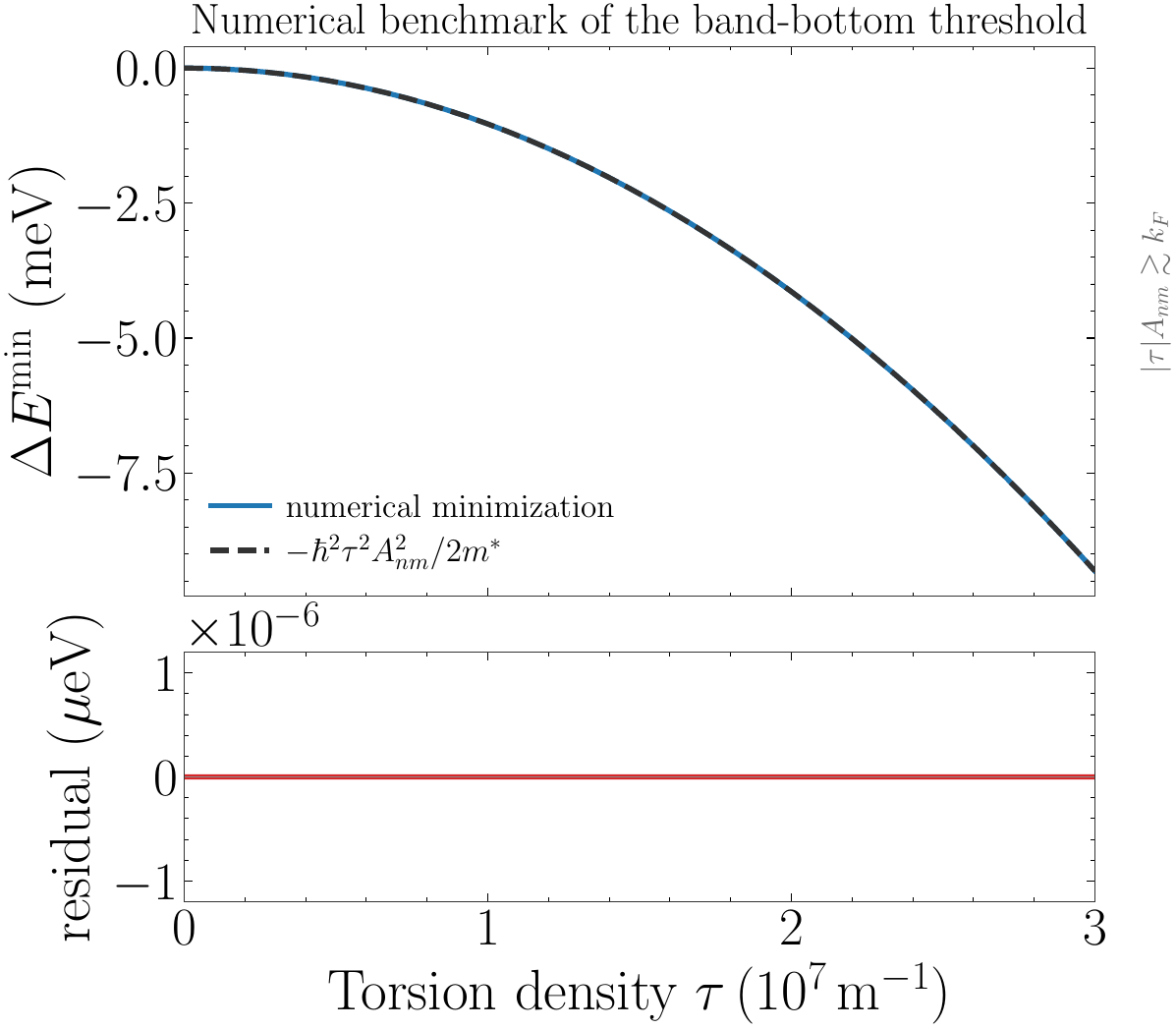}
  \caption{\textbf{Numerical consistency check of the band-bottom threshold.}
  The upper panel compares the band-bottom shift obtained by numerical minimization of the effective dispersion in Eq.~\eqref{eq:numerical_benchmark_dispersion} with the analytic result $-\hbar^2\tau^2A_{nm}^2/2m^*$. The two curves overlap on the meV scale, confirming that the Landauer channel-opening threshold is the subband minimum, not the fixed-$k_F$ transverse energy used for vertical optical transitions. The lower panel shows the residual $E^{\mathrm{min,num}}-E^{\mathrm{min,an}}$, which remains at numerical precision. The marker near the right edge indicates the approach to the perturbative bound $|\tau|A_{nm}\lesssim k_F$; beyond that scale, the effective parabolic lowering should be regarded as a parametric extrapolation rather than a controlled prediction.}
  \label{fig:numerical_benchmark}
\end{figure}

Figure~\ref{fig:numerical_benchmark} shows the outcome of this procedure for the representative mode used in the threshold and conductance figures. The numerical minimization and the analytic expression are indistinguishable on the scale of the band-bottom shift, while the residual remains at numerical precision. This agreement does not constitute a microscopic solution of the finite-radius nanowire: it does not include radial boundary eigenfunctions, spin-orbit mixing, self-consistent electrostatics, or contact-dependent scattering. Its role is more specific. It verifies that, within the effective-mass torsional Hamiltonian used in this work, the transport threshold is the true subband minimum and therefore decreases quadratically with torsion. The finite-radius and surface terms discussed in Sec.~\ref{sec:finite_edge_effects} should be understood as additional shifts and broadenings of this effective band bottom, whereas the fixed-$k_F$ energy $\varepsilon^{\mathrm{ch}}(\tau;k_F)$ remains the appropriate quantity for the vertical optical transition.

\section{Landauer--B\"uttiker transport formalism}
\label{sec:landauer_formalism}

To describe the electronic, thermal, and noise transport properties in the ballistic regime, we employ the Landauer--B\"uttiker scattering formalism~\cite{Landauer1957,Buttiker1986,Beenakker1997}. In this model, the fundamental quantity is the total transmission function $\Tcal(E)$, given by the sum of transmission probabilities for all energetically accessible transverse subbands. For the spin-resolved problem, it is useful to write
\begin{equation}
    \Tcal_\sigma(E) = \sum_{n,m} T_{n,m,\sigma}(E),
    \qquad
    \Tcal(E)=\sum_\sigma \Tcal_\sigma(E),
    \label{eq:Tspin}
\end{equation}
where $T_{n,m,\sigma}(E)$ incorporates the band-bottom channel threshold of Eq.~\eqref{eq:true_minimum}, contact transmission, longitudinal interference effects, and possible surface-roughness scattering. In the ideal adiabatic limit, $T_{n,m,\sigma}(E)$ approaches a step function when the Fermi energy crosses the corresponding band bottom. This is the same physical mechanism underlying quantized conductance in ballistic quantum point contacts~\cite{VanWees1988,Wharam1988}; in the present geometry, however, the subband bottom is mechanically displaced by torsion. A convenient form used in the numerical calculations is
\begin{equation}
    T^{(0)}_{n,m,\sigma}(E;\tau)
    =\frac{1}{1+\exp\left[(\Emin_{n,m,\sigma}(\tau)-E)/\Geff\right]},
    \label{eq:T0_smooth}
\end{equation}
with the replacement $\Emin\rightarrow\Emin{}^{,\edge}$ when finite-radius boundary corrections are included. We emphasize that the channel-opening condition encoded in Eq.~\eqref{eq:T0_smooth} is the propagation condition $E>\Emin_{n,m,\sigma}$ (equivalently $\mathcal{D}_\nu>0$), and not the value of the dispersion at the reservoir momentum $\varepsilon^{\mathrm{ch}}$, which is reserved for the optical resonance of Sec.~\ref{sec:optics}.

In the linear response regime, the electrical conductance $G$ and the Seebeck coefficient $S$ are determined by the moments of the transmission function weighted by the thermal derivative of the Fermi--Dirac distribution $f(E,\mu,T)$. We define the transport coefficients
\begin{equation}
    L_n(\mu,T) = \frac{1}{h} \sum_\sigma \int_{-\infty}^{\infty} (E-\mu)^n \left(-\frac{\partial f}{\partial E}\right) \Tcal_\sigma(E)\,dE.
    \label{eq:Ln_integrals}
\end{equation}
The factor of two commonly appearing in spin-degenerate Landauer formulas is replaced here by the explicit sum over $\sigma$, since the Zeeman term separates the two spin sectors. The electrical conductance and thermopower are then
\begin{equation}
    G = e^2 L_0,
    \qquad
    S = -\frac{1}{eT}\frac{L_1}{L_0}.
    \label{eq:G_S_defs}
\end{equation}
For spin caloritronic quantities, the same moments can be evaluated separately for the low- and high-energy Zeeman branches.

Beyond average currents, fluctuations in current provide information about charge granularity and transport coherence. Following the theory of Blanter and B\"uttiker~\cite{Blanter2000}, the zero-frequency shot-noise power density for non-interacting fermions can be written as
\begin{equation}
    S_{\mathrm{noise}} = 2eV\frac{e^2}{h}\sum_{n,m,\sigma} T_{n,m,\sigma}(1-T_{n,m,\sigma}),
\end{equation}
for a small applied bias $V$. The Fano factor, which quantifies the deviation from classical Poissonian noise, is
\begin{equation}
    F = \frac{S_{\mathrm{noise}}}{2eI}
      = \frac{\sum_{n,m,\sigma}T_{n,m,\sigma}(1-T_{n,m,\sigma})}
             {\sum_{n,m,\sigma}T_{n,m,\sigma}}.
    \label{eq:Fano_def}
\end{equation}
This set of equations constitutes the numerical basis for the results of quantized conductance, spin-resolved conductance imbalance, thermopower, and noise presented in this work.

\section{Finite-size and surface-boundary effects}
\label{sec:finite_edge_effects}

The dispersion in Eq.~\eqref{eq:Enm_spin} and the band-bottom thresholds in Eq.~\eqref{eq:true_minimum} provide the natural starting point for a translationally invariant torsion-controlled channel contacted by reservoirs. A real nanowire device, however, is neither infinite in the longitudinal direction nor boundary-free in the radial direction. It has a finite torsionally deformed segment of length $L$, a finite radius $R$, contacts to external reservoirs, and a microscopic surface structure. In this section, we make explicit how these finite-size and surface-boundary effects can be incorporated without abandoning the open-system Landauer description used above.

The key physical distinction is between a \emph{finite open quantum wire} and a \emph{closed quantum dot}. In the transport geometry considered here, the torsioned nanowire is modeled as a finite scattering region connected to two macroscopic reservoirs,
\begin{equation}
\begin{gathered}
    \mathrm{left\ reservoir}\;--\;0<z<L\;\mathrm{torsioned\ segment}\\
    --\;\mathrm{right\ reservoir}.
\end{gathered}
\end{equation}
The longitudinal coordinate is therefore finite inside the active device region, but the electronic states remain scattering states injected from the reservoirs. This is precisely the setting of the Landauer--B\"uttiker formalism~\cite{Landauer1957,Buttiker1986,Beenakker1997}. It differs from a closed box, where hard walls at $z=0$ and $z=L$ would quantize the longitudinal wave vector as $k_z\rightarrow q\pi/L$ and turn the device into an elongated quantum dot. Such a closed-dot limit is interesting, but it would require a resonant-level description with contact broadenings $\Gamma_L$ and $\Gamma_R$ rather than the simple open-channel conductance picture used here.

\subsection{Finite longitudinal length and Fabry--Perot interference}
\label{subsec:finite_length}

When the torsion is applied only over a finite interval $0<z<L$, the interfaces between the torsionally deformed segment and the untorsioned leads generally produce partial reflection. Even if the contacts are highly transparent, a small impedance mismatch between the propagating modes in the leads and the torsion-renormalized modes in the active region can generate longitudinal interference. The result is a Fabry--Perot modulation of the channel transmission, analogous to phase-coherent resonances observed in ballistic one-dimensional conductors such as carbon nanotube electron waveguides~\cite{Liang2001}.

For a single spin-resolved subband $\nu\equiv(n,m,\sigma)$, a minimal coherent model writes the transmission as
\begin{equation}
    T^{\FP}_{\nu}(E;\tau,L)
    = \frac{T_{L,\nu}T_{R,\nu}}
    {1+R_{L,\nu}R_{R,\nu}-2\sqrt{R_{L,\nu}R_{R,\nu}}\cos\Phi_\nu},
    \label{eq:TFP}
\end{equation}
where $T_{j,\nu}=1-R_{j,\nu}$ are the interface transparencies at the left and right contacts, and
\begin{equation}
    \Phi_\nu(E;\tau,L)=2k_\nu(E,\tau)L+\varphi_{L,\nu}+\varphi_{R,\nu}
    \label{eq:FPphase}
\end{equation}
contains the accumulated longitudinal phase. The wave vector $k_\nu(E,\tau)$ is obtained from the torsion-dependent dispersion relation, Eq.~\eqref{eq:Enm_spin}. Using Eq.~\eqref{eq:completed_square}, the two local roots are
\begin{align}
    k_{\nu,\pm}(E,\tau)
    &=-\tau A_{nm}\pm\sqrt{\frac{2m^*}{\hbar^2}\,\mathcal{D}_{\nu}(E,\tau)},
    \label{eq:k_roots}\\
    \mathcal{D}_{\nu}(E,\tau)
    &=E-\hbar\omega_cA_{nm}-\frac{\sigma}{2}\DeltaZ
      +\frac{\hbar^2\tau^2A_{nm}^2}{2m^*},
      \notag
\end{align}
when $\mathcal{D}_{\nu}(E,\tau)$ is positive. Note that $\mathcal{D}_\nu(E,\tau)=E-\Emin_{n,m,\sigma}(\tau)$, so the existence of real roots is exactly the channel-opening condition $E>\Emin_{n,m,\sigma}$ used in Sec.~\ref{subsec:kz_thresholds}. The sign is chosen so that the group velocity points in the direction of propagation. The phases $\varphi_{L,\nu}$ and $\varphi_{R,\nu}$ describe reflection phases at the two contacts. Thus, both the Fabry--Perot phase and the channel-opening condition in the low-bias conductance are governed by the band bottom of Eq.~\eqref{eq:true_minimum}.

Equation~\eqref{eq:TFP} has two important limiting cases. In the adiabatic-contact limit $R_{L,\nu},R_{R,\nu}\ll1$, the Fabry--Perot oscillations are weak and one recovers smooth conductance plateaus. For stronger mismatch, the same subband thresholds remain present, but the conductance develops oscillatory resonances as a function of $E_F$, $B$, $\tau$, or $L$. Thus, finite length does not invalidate the torsion-control mechanism; it dresses the ideal staircase by a coherent interference envelope.

In realistic devices, phase coherence and elastic boundary scattering reduce the visibility of the oscillations. A useful phenomenological form is
\begin{align}
    T_\nu(E;\tau,L)
    &=\overline{T}_\nu(E;\tau)
      +e^{-L/L_\phi}e^{-L/\ell_{\edge,\nu}}\notag\\ &\times
       \Delta T^{\FP}_\nu(E;\tau,L),
       \label{eq:dampedFP}\\
    \Delta T^{\FP}_\nu(E;\tau,L)
    &=T^{\FP}_\nu(E;\tau,L)-\overline{T}_\nu(E;\tau).
      \notag
\end{align}
where $L_\phi$ is the phase-coherence length, $\ell_{\edge,\nu}$ is a surface-limited elastic mean free path, and $\overline{T}_\nu$ is the incoherent or phase-averaged transmission. In the short, clean limit $L\ll L_\phi,\ell_{\edge,\nu}$, coherent oscillations are visible; in the long or rough limit, the oscillatory term is exponentially suppressed and only the broadened subband thresholds remain.

We stress that, within the controlled regime of Eq.~\eqref{eq:validity_hierarchy}, the phase-averaged ideal conductance is a stable, single-valued function of the torsion magnitude. Since the ideal band-bottom contribution decreases as $-\tau^2$ and each channel transmission is bounded in $[0,1]$, the phase-averaged open-channel conductance is monotonically non-decreasing in $|\tau|$ when the surface correction is a smooth common threshold shift. This statement should not be interpreted as a general theorem for arbitrary microscopic boundaries: a strongly $\tau$-dependent surface potential or contact mismatch can add reversible structure on top of the staircase. In the present noninteracting, linear-response model, however, there is no feedback mechanism capable of producing bistability or hysteresis in $G(\tau)$. The only departures from a smooth staircase are the coherent Fabry--Perot oscillations of Eq.~\eqref{eq:TFP}, which are reversible and average out under decoherence, and the shaded region $|\tau|A_{nm}\gtrsim k_F$, where the parabolic band-bottom lowering ceases to be controlled. Genuine instability would require ingredients deliberately omitted here, self-consistent electrostatics, in which the charge of newly opened channels screens the confining potential and feeds back on the thresholds, or a dynamical torsion treated as a mechanical degree of freedom rather than an external control parameter, for which electronic back-action on the torsional mode could in principle drive a torsiomechanical instability. Both lie beyond the present scope and would be natural directions for a self-consistent or optomechanical extension.

\subsection{Finite radial radius and boundary conditions}
\label{subsec:radial_boundary}

The ideal torsion-induced radial confinement should also be supplemented by the physical surface of the nanowire. Let $R$ be the mean radius of the wire. The radial wave function $\chi_{n,m}(r)$ must satisfy a boundary condition at $r=R$. The simplest choice is a hard-wall condition,
\begin{equation}
    \chi_{n,m}(R)=0,
    \label{eq:dirichlet}
\end{equation}
which represents an infinite surface barrier. A more flexible and experimentally realistic description is a mixed or Robin boundary condition,
\begin{equation}
    \left.\left(\partial_r+\lambda_s^{-1}\right)\chi_{n,m}(r)\right|_{r=R}=0,
    \label{eq:robin}
\end{equation}
where $\lambda_s$ is an effective surface extrapolation length. Robin conditions are widely used as effective long-wavelength descriptions of short-range surface potentials and finite interface barriers~\cite{BelchevWalton2010}. In the present context, $\lambda_s\rightarrow0$ approaches a hard-wall-like limit, whereas $|\lambda_s|\rightarrow\infty$ approaches a Neumann-like boundary.

The finite-radius spectrum can be written as
\begin{align}
    E^{\edge}_{n,m,\sigma}(k_z)
    &= E^{(0)}_{n,m,\sigma}(k_z)
    + \delta\varepsilon_{n,m}(R,\lambda_s,\tau).
    \label{eq:Eedge}
\end{align}
where $E^{(0)}_{n,m,\sigma}$ denotes the torsion-plus-Zeeman spectrum in Eq.~\eqref{eq:Enm_spin}. The correction $\delta\varepsilon_{n,m}$ is obtained by solving the radial eigenvalue problem with Eq.~\eqref{eq:dirichlet} or Eq.~\eqref{eq:robin}. In the weak-torsion or large-radius limit, this correction reduces to the familiar transverse quantization scale of a cylindrical quantum wire,
\begin{equation}
    \delta\varepsilon_{n,m}^{\mathrm{hw}}(R)
    \simeq \frac{\hbar^2\alpha_{|m|,n}^2}{2m^*R^2},
    \label{eq:hardwallshift}
\end{equation}
where $\alpha_{|m|,n}$ is the $n$th zero of the Bessel function $J_{|m|}$. In the torsion-dominated regime, Eq.~\eqref{eq:hardwallshift} should be viewed only as a scale estimate, since the radial eigenfunctions are those of the effective torsional confinement rather than pure Bessel modes. Nevertheless, the qualitative role of the boundary is clear: a smaller radius increases the subband thresholds, whereas a soft surface potential or finite penetration length shifts them more weakly.

This finite-radius correction modifies the spin-dependent channel thresholds according to Eq.~\eqref{eq:lead_threshold_edge}. Since the boundary correction is mostly orbital, it shifts both spin projections in a similar way, while the relative spin offset continues to be controlled by the Zeeman energy $\DeltaZ=|g^*|\mu_BB$. Therefore, finite radius changes the torsion value at which a given channel opens or closes, but it does not by itself remove the spin-selective interval unless the surface-induced broadening becomes comparable to the Zeeman splitting. In terms of Eq.~\eqref{eq:tau_open}, the boundary correction mainly produces a common displacement of the two spin-resolved opening points, whereas their separation remains governed by Eq.~\eqref{eq:spin_window}.

\subsection{Roughness and surface-induced broadening}
\label{subsec:roughness}

Real nanowire boundaries are not perfectly cylindrical. Surface disorder, etching imperfections, interface roughness, and radial radius fluctuations can be represented by
\begin{equation}
    R(z,\varphi)=R_0+\delta R(z,\varphi),
    \qquad
    \langle\delta R\rangle=0,
    \label{eq:Rrough}
\end{equation}
with a typical correlator such as
\begin{equation}
    \langle\delta R(z)\delta R(z')\rangle
    =\Delta_R^2\exp\left[-\frac{(z-z')^2}{\Lambda_R^2}\right],
    \label{eq:roughcorr}
\end{equation}
where $\Delta_R$ is the root-mean-square roughness amplitude and $\Lambda_R$ is the correlation length. Rough boundaries in quasi-one-dimensional wires are known to produce backscattering, mode mixing, and a crossover from quasi-ballistic to diffusive or localized transport~\cite{Feilhauer2011}. In semiconductor nanowires, interface roughness can strongly modify the mobility and may even induce ballistic-to-diffusive crossovers depending on radius, density, and length~\cite{Fu2018}.

For the present model, the leading effect of roughness can be captured by a mode-dependent broadening of the subband onset,
\begin{equation}
    \Gamma_{\edge,\nu}(E)
    \sim \frac{\hbar v_\nu(E)}{2\ell_{\edge,\nu}(E)},
    \label{eq:Gamma_edge}
\end{equation}
where $v_\nu(E)=\hbar^{-1}\partial E_\nu/\partial k_z$ is the longitudinal group velocity and $\ell_{\edge,\nu}$ is the surface-limited elastic mean free path. A convenient numerical implementation is to replace the ideal sharp threshold with a broadened activation function,
\begin{equation}
    \Theta(E-\Emin_\nu)
    \longrightarrow
    \frac{1}{1+\exp\left[(\Emin{}^{,\edge}_\nu-E)/\Gamma_{\edge,\nu}\right]}.
    \label{eq:smooth_threshold}
\end{equation}
Combining this threshold with Eq.~\eqref{eq:TFP} gives a transmission of the form
\begin{align}
    T_{\nu}^{\edge}(E;\tau,L,R)
    &= \frac{1}{1+\exp\left[(\Emin{}^{,\edge}_\nu-E)/\Gamma_{\edge,\nu}\right]}\notag\\&\times T_{\nu}^{\FP}(E;\tau,L).
    \label{eq:Tedge}
\end{align}
This phenomenological transmission is not a substitute for a full microscopic scattering-matrix calculation. It provides a compact parametrization of the crossover from an ideal ballistic channel to a realistic nanowire with finite surface quality. The spin-selective effect survives when
\begin{equation}
    \Geff \ll \DeltaZ
    \label{eq:robust_condition}
\end{equation}
and when the active segment remains shorter than the localization length. Under these conditions, surface roughness smooths the conductance steps and reduces the maximum polarization, but it does not close the torsion-tunable spin-selective interval. This is the physical content of the contrast estimate in Eq.~\eqref{eq:contrast_estimate}.

\subsection{Consequences for optical transitions}
\label{subsec:optical_edge}

Finite-radius boundary conditions also affect the optical response. The electric-dipole matrix element between two eigenstates is
\begin{equation}
    M_{fi}=\langle\psi_f|e\mathbf{E}\cdot\mathbf{r}|\psi_i\rangle,
    \label{eq:dipole_edge}
\end{equation}
where the wave functions now satisfy the radial boundary condition at $r=R$. For a transverse optical field, for example $\mathbf{E}\parallel\hat{x}$, one has $x=r\cos\varphi$, so the angular selection rule is
\begin{equation}
    \Delta m=\pm1,
    \label{eq:transverse_selection}
\end{equation}
while the radial overlap determines which transitions in $n$ carry the strongest oscillator strength. In the idealized harmonic-like torsional confinement, the dominant low-energy oscillator strength is often associated with neighboring radial branches. With finite boundaries, however, the more precise statement is that the resonance is governed by the full finite-radius energy difference and by the dipole matrix element in Eq.~\eqref{eq:dipole_edge}.

For a spin-conserving electric-dipole transition, the optical resonance therefore becomes
\begin{align}
    \hbar\omega^{\edge}_{fi}(\tau,R,L)
    &= E^{\edge}_{f}(k_z)-E^{\edge}_{i}(k_z) \notag\\
    &= \hbar(\wt+\wc)\left(A_f-A_i\right)
    +\delta\varepsilon_f(R,\lambda_s,\tau)
    \notag\\&-\delta\varepsilon_i(R,\lambda_s,\tau),
    \label{eq:omega_edge}
\end{align}
for the corresponding pair of transverse branches. The transition is vertical, so both states are evaluated at the same longitudinal momentum $k_z$; in the open device, $k_z$ is fixed by the energy of the incident carriers and by the dispersion relation, $k_z\simeq\kF$, which is the physical origin of the fixed-momentum quantity $\varepsilon^{\mathrm{ch}}$ of Eq.~\eqref{eq:lead_threshold}. Only in the closed-dot limit would one replace $k_z$ by $q\pi/L$, producing an explicitly length-quantized optical ladder,
\begin{equation}
    \omega^{(q)}_{fi}(\tau,L)
    = \left(\frac{\hbar q\pi}{m^*L}\tau + \frac{eB}{m^*}\right)
      \left(A_f-A_i\right),
    \label{eq:closed_length_optics}
\end{equation}
which is not the primary transport geometry considered here. Thus, in the experimentally relevant open device, finite $L$ mainly controls longitudinal interference and linewidth, whereas finite $R$ and surface quality shift and broaden the transverse optical resonances.

\section{Spin-resolved conductance and robustness}
\label{sec:spin_conductance}

The threshold picture in Fig.~\ref{fig:threshold_mechanism} can be tested directly in the spin-resolved conductance. At low bias, each spin branch contributes according to
\begin{equation}
    G_\sigma=e^2L_{0,\sigma},
    \qquad
    L_{0,\sigma}=\frac{1}{h}\int dE\left(-\frac{\partial f}{\partial E}\right)\mathcal{T}_\sigma(E),
\end{equation}
with the channel transmission controlled by Eq.~\eqref{eq:T0_smooth}. The current polarization is defined as
\begin{equation}
    P=\frac{G_{\mathrm{low}}-G_{\mathrm{high}}}
    {G_{\mathrm{low}}+G_{\mathrm{high}}},
    \label{eq:polarization_low_high}
\end{equation}
where the labels refer to the lower- and higher-energy Zeeman branches. This convention is equivalent to a spin polarization once the sign of $g^*$ and the quantization axis are fixed.

\begin{figure}[t]
  \centering
  \includegraphics[width=0.98\linewidth]{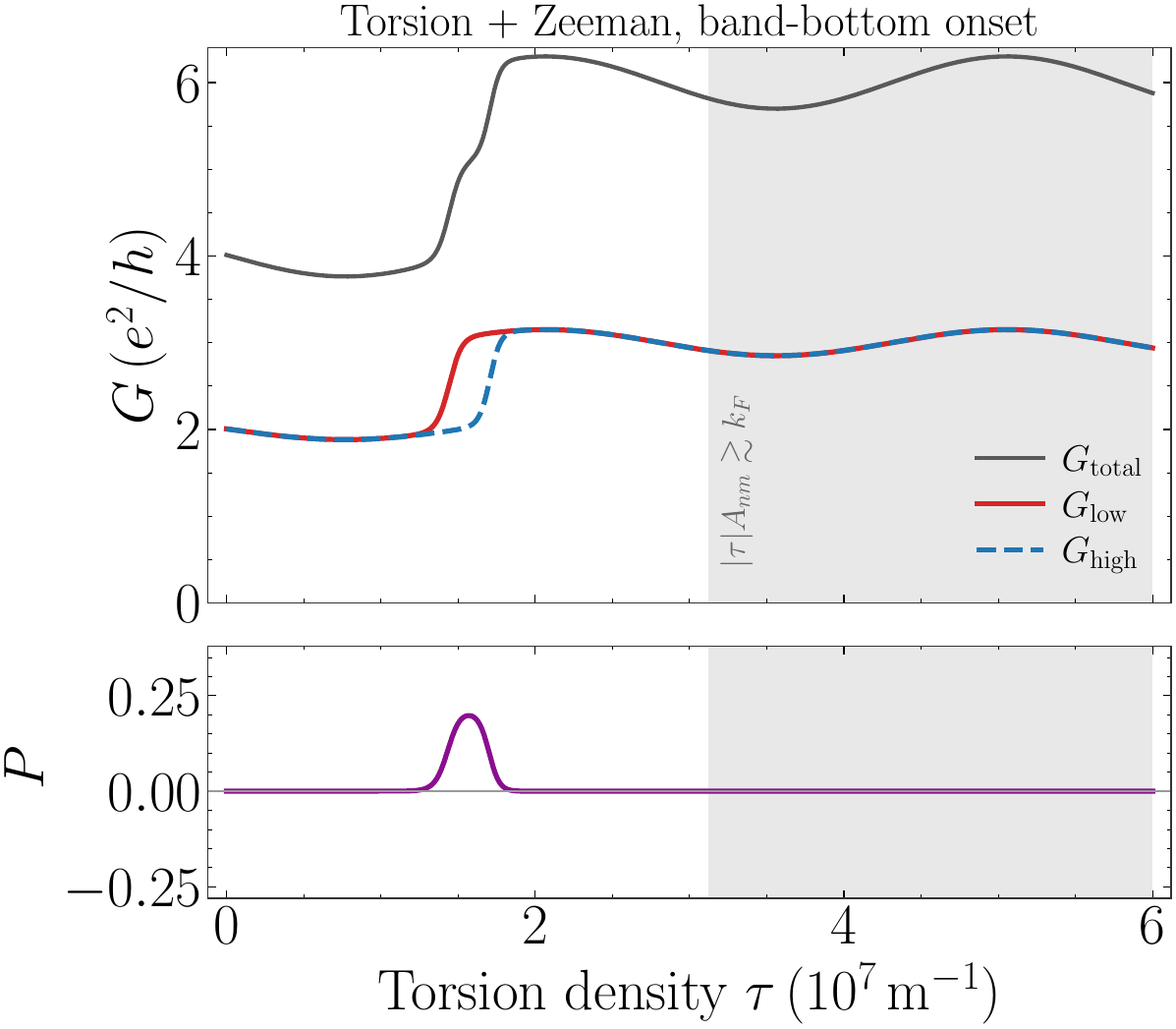}
  \caption{\textbf{Spin-resolved conductance imbalance driven by torsion.}
  Representative conductance curves for the two Zeeman branches and the total conductance as a function of torsion density, computed with the band-bottom onset of Eq.~\eqref{eq:true_minimum}: torsion lowers each subband bottom, so channels open and the conductance rises with $\tau$. The separation between the branch onsets follows the scale shown in Fig.~\ref{fig:threshold_mechanism}. Finite temperature and surface-induced broadening smooth the conductance steps, but a finite polarization remains whenever the two onsets are resolved on the scale $\Gamma_{\mathrm{eff}}$. The shaded region marks $|\tau|A_{nm}\gtrsim k_F$, where the parabolic band-bottom lowering ceases to be controlled [Eq.~\eqref{eq:kF_condition}].}
  \label{fig:Spin_Filter}
\end{figure}

Figure~\ref{fig:Spin_Filter} illustrates the basic transport effect. As $\tau$ is swept, the band bottoms of the two Zeeman branches cross $E_F$ at different values. In the intermediate region, the threshold mode of the lower-energy branch has opened while the corresponding mode of the higher-energy branch remains suppressed, producing a spin-selective conductance imbalance on top of the background conductance. The result is not tied to a perfectly sharp step function: the same mechanism remains operative for broadened onsets, provided the hierarchy in Eq.~\eqref{eq:robust_condition} is satisfied. We note that the maximum polarization attained in Fig.~\ref{fig:Spin_Filter} is moderate because the spin imbalance develops on top of a background of already-open subbands; the large branch contrast quantified in Sec.~\ref{sec:transistor} is instead reached in the activated regime, where both absolute conductances are small. The device therefore operates either as a high-contrast, low-current spin filter or as a moderate-polarization, appreciable-current channel, but not as both simultaneously.

\begin{figure}[t]
  \centering
  \includegraphics[width=0.98\linewidth]{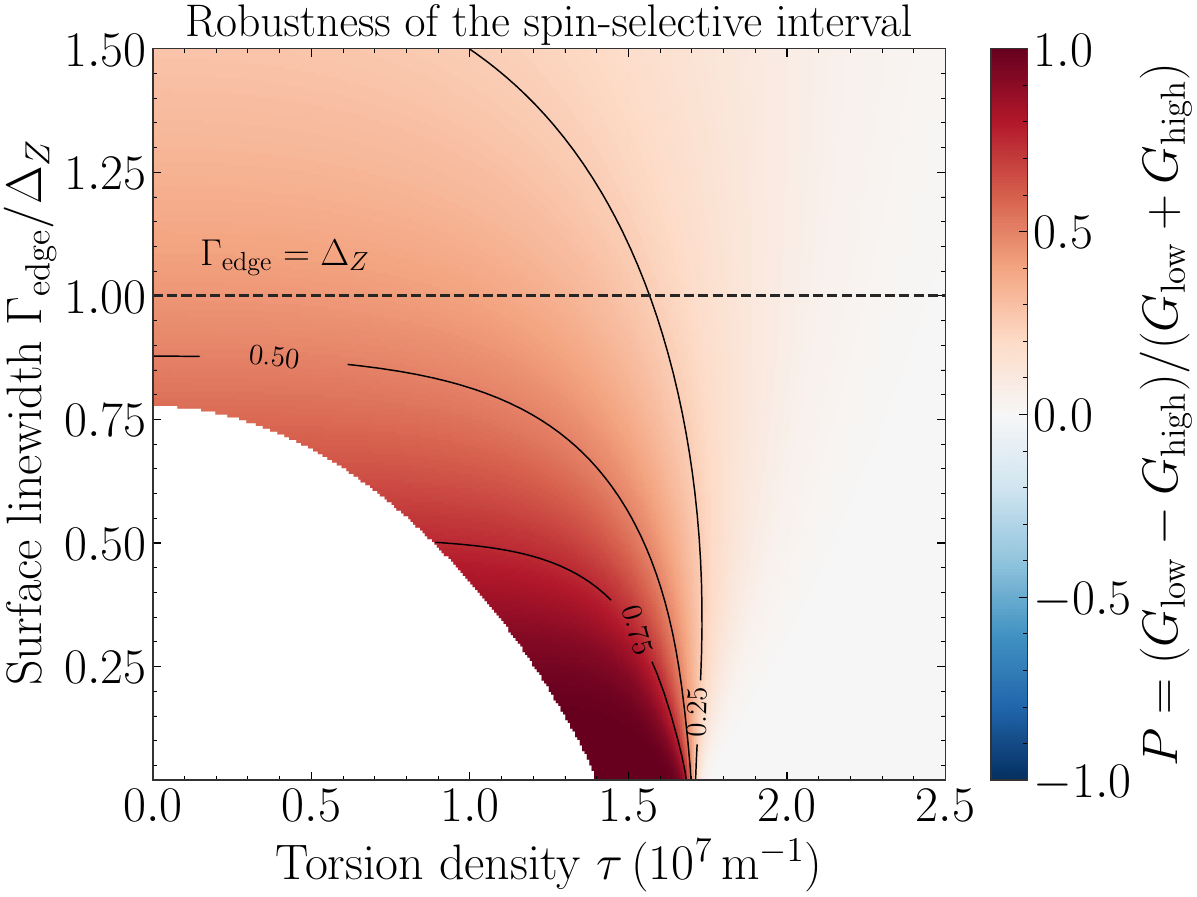}
  \caption{\textbf{Robustness of the spin-selective interval.}
  Polarization map $P=(G_{\mathrm{low}}-G_{\mathrm{high}})/(G_{\mathrm{low}}+G_{\mathrm{high}})$ as a function of torsion density and normalized surface linewidth $\Gamma_{\mathrm{edge}}/\Delta_Z$. The dashed line marks $\Gamma_{\mathrm{edge}}=\Delta_Z$. The high-polarization region survives below this scale and is progressively washed out when surface-induced linewidths become comparable to or exceed the Zeeman splitting.}
  \label{fig:robustness_map}
\end{figure}

The robustness criterion is quantified in Fig.~\ref{fig:robustness_map}. The relevant control parameter is not the absolute magnitude of the surface disorder, but its linewidth measured relative to the Zeeman energy. For $\Gamma_{\mathrm{edge}}/\Delta_Z<1$, sizeable polarization persists over a finite torsion interval. For larger broadening, the two spin-resolved onsets overlap, and the polarization decreases. This map is the numerical counterpart of the analytic estimate in Eq.~\eqref{eq:contrast_estimate}.

\begin{figure}[t]
  \centering
  \includegraphics[width=0.98\linewidth]{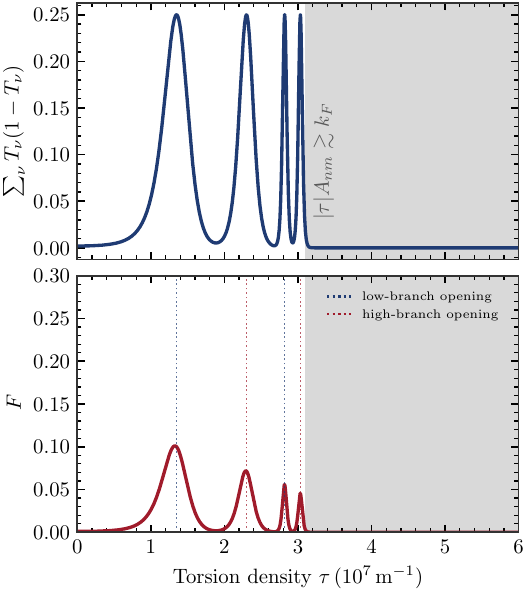}
  \caption{\textbf{Shot noise and Fano factor across torsion-driven channel openings.}
  Upper panel: normalized zero-frequency shot-noise power $\sum_\nu T_\nu(1-T_\nu)$ [Eq.~\eqref{eq:Fano_def}] as a function of torsion density. Lower panel: Fano factor $F=\sum_\nu T_\nu(1-T_\nu)/\sum_\nu T_\nu$. Transmissions are evaluated at $E=\EF$ using the band-bottom onset of Eq.~\eqref{eq:T0_smooth}. Each orbital threshold appears as a pair of peaks, one per Zeeman branch; the in-pair splitting reproduces the spin-selective interval $\Delta\tau_{n,m}$ of Eq.~\eqref{eq:spin_window}, while $F\to0$ on the intervening plateaus where channels are fully transmitting. The shaded region marks $|\tau|A_{nm}\gtrsim k_F$, where the parabolic band-bottom lowering ceases to be controlled [Eq.~\eqref{eq:kF_condition}].}
  \label{fig:fano}
\end{figure}

The same channel structure determines the current fluctuations of Eqs.~\eqref{eq:Fano_def}. Figure~\ref{fig:fano} shows the normalized zero-frequency shot-noise power $\sum_\nu T_\nu(1-T_\nu)$ and the Fano factor $F$ as functions of torsion density, computed with the band-bottom transmissions of Eq.~\eqref{eq:T0_smooth} evaluated at $E=\EF$. On a conductance plateau, every open channel has $T_\nu\to1$, and the partition noise is suppressed ($F\to0$); at each torsion-driven channel opening, one mode passes through at $T_\nu\simeq1/2$, and both the shot-noise power and $F$ exhibit a local maximum. Because the two Zeeman branches of a given orbital mode open at the two torsion values $\tau^{\mathrm{open}}_{n,m,\sigma}$ of Eq.~\eqref{eq:tau_open}, each orbital threshold produces a \emph{pair} of closely spaced Fano peaks whose splitting reproduces the spin-selective interval $\Delta\tau_{n,m}$ of Eq.~\eqref{eq:spin_window}. The shot noise therefore provides an independent, spin-resolved fingerprint of the torsion-controlled thresholds, complementary to the conductance imbalance of Fig.~\ref{fig:Spin_Filter}: the peak \emph{positions} locate the channel openings, while the peak \emph{splitting} measures the Zeeman-induced spin separation. As the surface-induced linewidth grows toward $\Delta_Z$, neighboring peaks in each pair merge, in line with the same robustness criterion $\Geff\ll\DeltaZ$ controlling the conductance and thermopower responses.

\section{Tunable intersubband optical absorption}
\label{sec:optics}

\begin{figure}[t]
  \centering
  \includegraphics[width=0.98\linewidth]{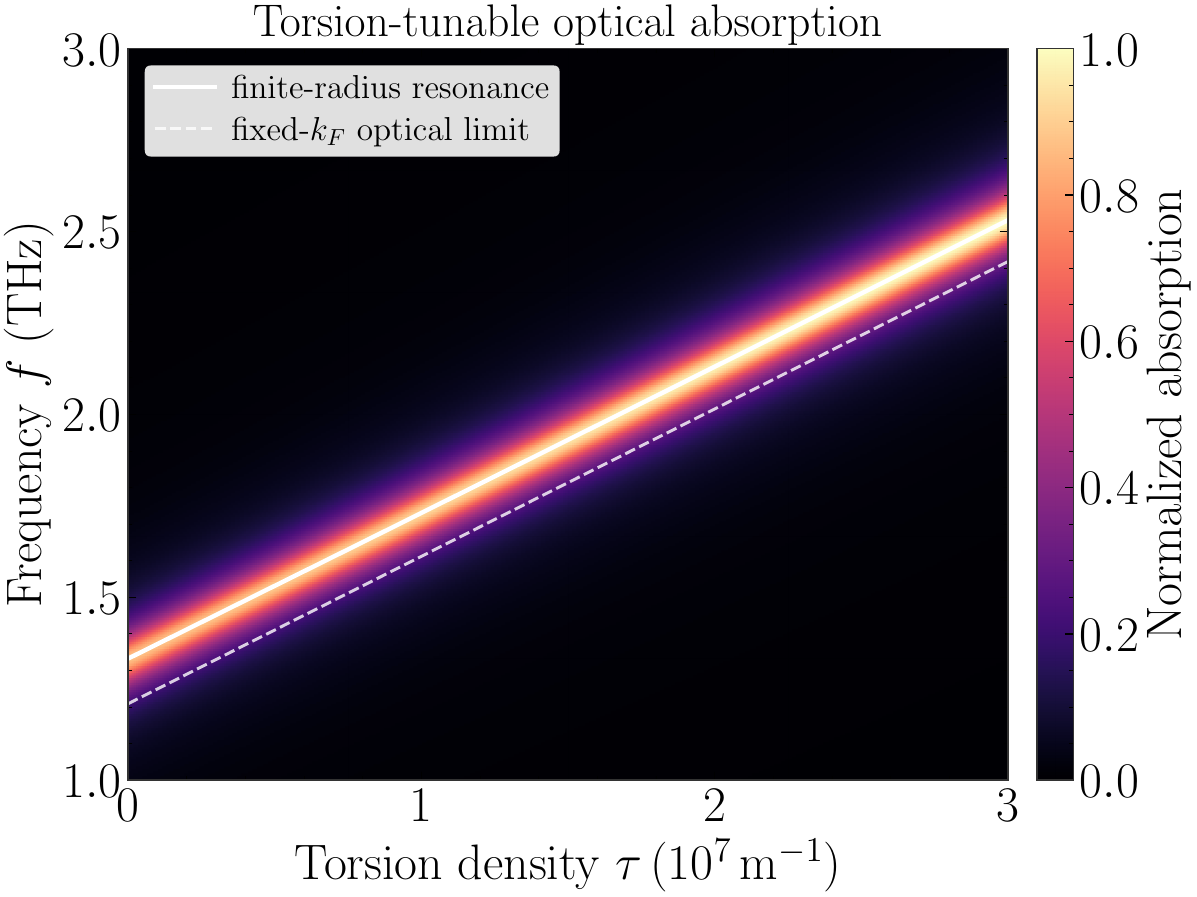}
  \caption{\textbf{Torsion-tunable optical absorption.}
  Normalized intersubband absorption in the $(\tau,f)$ plane. The bright ridge follows the finite-radius resonance, while the dashed curve shows the ideal fixed-$k_F$ optical limit. Radial boundary conditions shift the absolute resonance frequency, but the leading torsional tunability remains approximately linear over the parameter range shown.}
  \label{fig:optical_absorption_map}
\end{figure}

We now consider a linearly polarized optical field perpendicular to the nanowire axis, $\mathbf{E}\perp\hat{z}$. Intersubband transitions are a standard route to infrared and THz light--matter coupling in semiconductor nanostructures~\cite{WestEglash1985,Faist1994,Williams2007,Tonouchi2007}. Optical absorption is a vertical (momentum-conserving) transition, so the relevant transverse energies are evaluated at the occupied longitudinal momentum $k_z\simeq\kF$; this is precisely the fixed-$k_z$ quantity $\varepsilon^{\mathrm{ch}}$ introduced in Eq.~\eqref{eq:lead_threshold}, and not the transport band bottom of Eq.~\eqref{eq:true_minimum}. The resonance frequency for a transition $i\rightarrow f$ is
\begin{equation}
    \hbar\omega_{fi}=\varepsilon^{\mathrm{ch}}_f(\tau;\kF)-
    \varepsilon^{\mathrm{ch}}_i(\tau;\kF).
    \label{eq:omega_fi_ch}
\end{equation}
For neighboring torsion-renormalized branches with $A_f-A_i=1$, the $\tau^2$ contributions cancel in the difference and one obtains the leading fixed-momentum optical relation
\begin{equation}
    \omega_{\mathrm{res}}(\tau)=\frac{\hbar\kF}{m^*}\tau+\frac{eB}{m^*}.
    \label{eq:omega_res_step3}
\end{equation}
Finite radius and surface boundary conditions modify this through Eq.~\eqref{eq:omega_edge}. We stress that Eq.~\eqref{eq:omega_res_step3} captures only the torsional and magnetic slopes of the resonance; it omits the finite-radius confinement contribution and should not be read as the absolute resonance frequency. In particular, the apparent vanishing of $\omega_{\mathrm{res}}$ as $\tau,B\to0$ is an artifact of this slope-only expression: the physical transverse spacing of a confined wire remains finite and is restored by the boundary term $\delta\varepsilon_f-\delta\varepsilon_i$ in Eq.~\eqref{eq:omega_edge} [cf. Eq.~\eqref{eq:hardwallshift}]. For the representative radii of Table~\ref{tab:parameters}, this confinement offset is itself in the THz range [e.g.\ for $R=50~\mathrm{nm}$, Eq.~\eqref{eq:hardwallshift} gives a lowest transverse spacing of order $1~\mathrm{meV}$, i.e.\ a few hundred GHz], so that the absolute resonance of Fig.~\ref{fig:optical_absorption_map} sits on a finite zero-torsion offset and is tuned upward by torsion. The optical activity is determined by the dipole matrix element $M_{fi}$ in Eq.~\eqref{eq:dipole_edge}; for transverse polarization, the angular selection rule is $\Delta m=\pm1$.

Because $\omega_\tau=\hbar k_z\tau/m^*$ is itself momentum dependent, the same mechanism that makes the resonance torsion-tunable also implies that, when the absorption is summed over the occupied longitudinal states $0\le k_z\lesssim\kF$, the line acquires an inhomogeneous torsional broadening of order $\hbar\tau v_F$ in addition to the intrinsic width $\Gamma_{\mathrm{opt}}$. To display both the resonance energy and the oscillator-strength broadening, we use a Lorentzian absorption model,
\begin{equation}
    \mathcal{A}(\omega,\tau)\propto
    \sum_{i,f}(f_i-f_f)|M_{fi}|^2
    \frac{\Gamma_{\mathrm{opt}}}
    {(\hbar\omega-E_f+E_i)^2+\Gamma_{\mathrm{opt}}^2},
    \label{eq:absorption_lorentzian}
\end{equation}
where $\Gamma_{\mathrm{opt}}$ is the optical linewidth. This minimal absorption model resolves the resonance position and linewidth; a fully momentum-resolved evaluation, integrating Eq.~\eqref{eq:absorption_lorentzian} over the occupied $k_z$, would additionally capture the torsional broadening noted above. A microscopic evaluation of $|M_{fi}|^2$ with finite-radius eigenfunctions would refine the intensity distribution but would not change the leading torsional slope.

Figure~\ref{fig:optical_absorption_map} shows the resulting optical response. Instead of plotting only an energy difference, the map displays the resonance as an optically active absorption ridge. The comparison between the finite-radius ridge and the fixed-$k_F$ optical limit makes clear that the surface mainly produces a static energy offset and a finite linewidth, while the derivative $\partial\omega/\partial\tau\simeq\hbar\kF/m^*$ is controlled by the torsion-coupled longitudinal momentum in the fixed-momentum optical regime. We note that this linear slope is precisely the contribution that cancels for the band-bottom transport threshold but survives for the fixed-$k_z$ vertical transition, which is why the optical and transport quantities use different threshold conventions.

\section{Spin caloritronics: spin thermopower protocol}
\label{sec:spin_seebeck}

The spin-resolved thresholds also generate a spin thermoelectric response. The sensitivity of thermopower to sharp energy-dependent transmission is a central result of mesoscopic thermoelectrics~\cite{Sivan1986,Butcher1990,HicksDresselhaus1993b,MahanSofo1996}. For each Zeeman branch we define the moments
\begin{equation}
    L_{n,\sigma}=\frac{1}{h}\int dE\,(E-\mu)^n
    \left(-\frac{\partial f}{\partial E}\right)\mathcal{T}_\sigma(E),
\end{equation}
so that the linear current carried by branch $\sigma$ can be written as
\begin{equation}
    I_\sigma=e^2L_{0,\sigma}\Delta V_\sigma
    +\frac{e}{T}L_{1,\sigma}\Delta T.
    \label{eq:spin_resolved_current}
\end{equation}
The branch thermopower is
\begin{equation}
    S_\sigma=-\frac{1}{eT}\frac{L_{1,\sigma}}{L_{0,\sigma}},
\end{equation}
and we define
\begin{equation}
    S_s=S_{\mathrm{low}}-S_{\mathrm{high}}.
    \label{eq:Sspin_step3}
\end{equation}
The corresponding charge and spin currents are $I_c=I_{\mathrm{low}}+I_{\mathrm{high}}$ and $I_s=(\hbar/2e)(I_{\mathrm{low}}-I_{\mathrm{high}})$. Equation~\eqref{eq:Sspin_step3} is therefore the thermopower difference associated with the two spin-resolved transport channels; experimentally, it corresponds to a spin accumulation under open-charge-current conditions or to detection with spin-selective contacts.

\begin{figure}[t]
  \centering
  \includegraphics[width=0.98\linewidth]{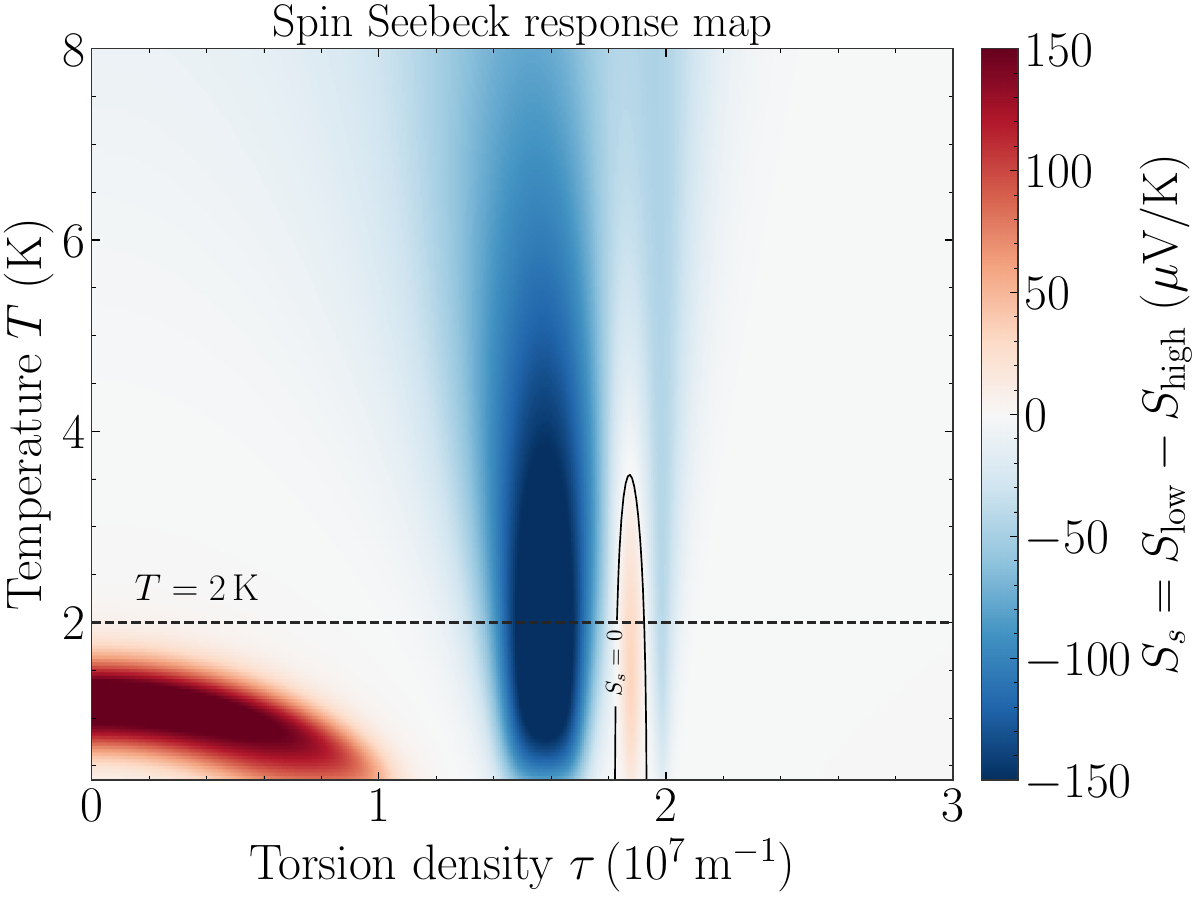}
  \caption{\textbf{Spin Seebeck response map.}
  Spin thermopower $S_s=S_{\mathrm{low}}-S_{\mathrm{high}}$ as a function of torsion density and temperature. The dashed line indicates $T=2~\mathrm{K}$, connecting the map with the low-temperature transport regime. Black contours mark sign reversals of $S_s$, showing that the thermally generated spin response can be inverted by changing torsion.}
  \label{fig:spin_seebeck_map}
\end{figure}

Figure~\ref{fig:spin_seebeck_map} shows that the spin thermopower changes sign as the spin-resolved band bottoms pass through the thermal window around $E_F$. Increasing temperature broadens the response and reduces fine threshold structure, while the sign-changing regions persist as long as the two branch onsets remain thermally distinguishable. Surface-induced broadening has an analogous effect and is governed by the same linewidth scale that enters Eq.~\eqref{eq:Gamma_eff}.

\section{Spin-selective transfer characteristic}
\label{sec:transistor}

The same threshold separation produces a transistor-like spin-selective transfer curve when torsion drives one Zeeman branch into the transmitting window while the other is still closed. We quantify this effect by the branch contrast
\begin{equation}
    \mathcal{R}_{\mathrm{spin}}=\frac{G_{\mathrm{low}}}{G_{\mathrm{high}}},
    \qquad
    \log_{10}\mathcal{R}_{\mathrm{spin}}
    =\log_{10}\left(\frac{G_{\mathrm{low}}}{G_{\mathrm{high}}}\right).
\end{equation}
This definition avoids implying a particular microscopic readout architecture; it simply measures how strongly torsion selects one spin-resolved transport branch over the other.
\begin{figure}[t]
  \centering
  \includegraphics[width=0.98\linewidth]{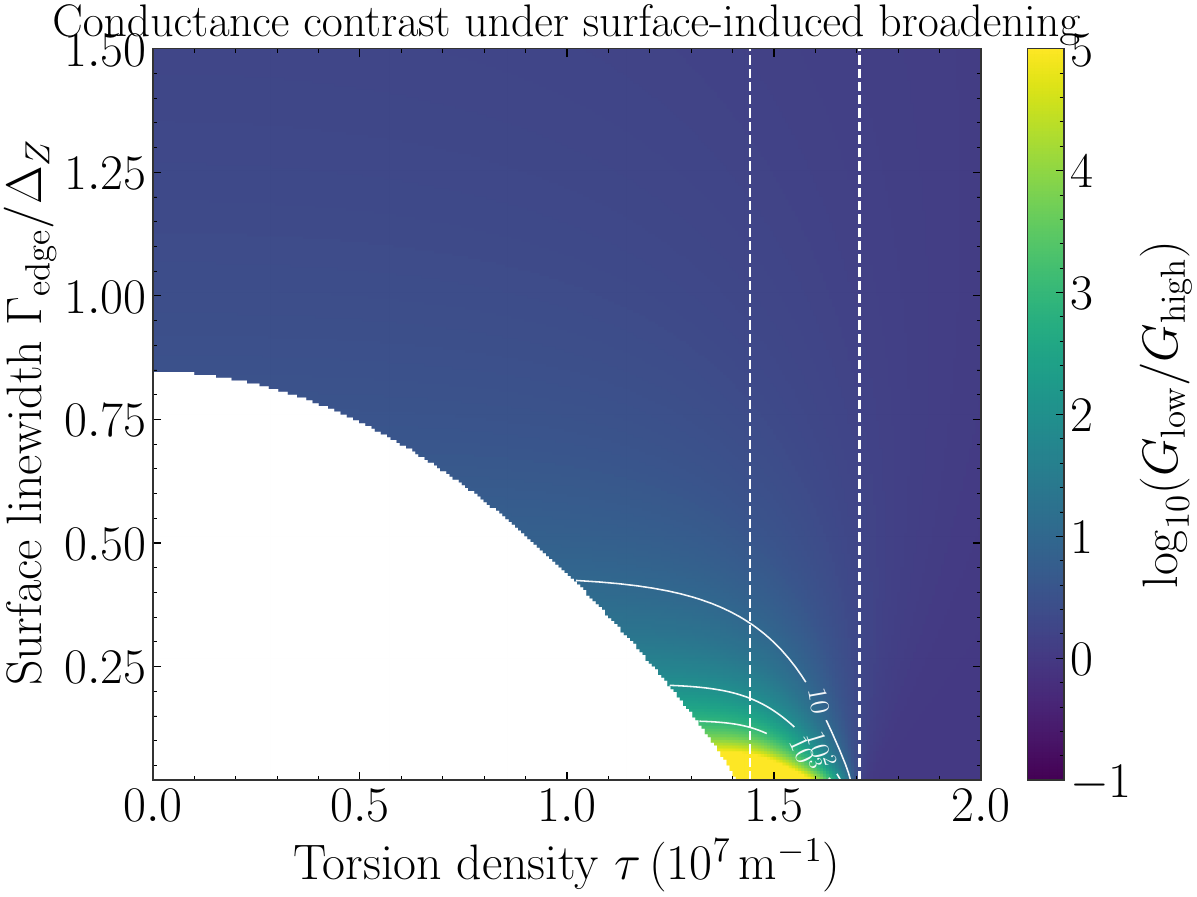}
  \caption{\textbf{Conductance contrast under surface-induced broadening.}
  Map of $\log_{10}(G_{\mathrm{low}}/G_{\mathrm{high}})$ as a function of torsion density and normalized surface linewidth. Dashed vertical lines indicate the spin-selective torsion interval inferred from the threshold mechanism. Contours mark contrasts of $10$, $10^2$, and $10^3$. Large branch contrast is obtained in the clean regime and is reduced continuously as $\Gamma_{\mathrm{edge}}$ approaches the Zeeman scale.}
  \label{fig:transistor_contrast_map}
\end{figure}

Figure~\ref{fig:transistor_contrast_map} summarizes the transfer behavior in the presence of surface-induced broadening. The high-contrast region lies inside the spin-selective torsion interval and is most pronounced for $\Gamma_{\mathrm{edge}}/\Delta_Z<1$. This result refines the device interpretation: the proposed structure should be viewed as a torsion-controlled spin-selective scatterer whose operation depends on the hierarchy $\Gamma_{\mathrm{eff}}\ll\Delta_Z$, rather than as an ideal disorder-independent switch.

\section{Conclusion}
\label{sec:conclusion}

We have developed a spin-resolved open-channel theory for a semiconductor nanowire with screw-dislocation-induced torsion. The main result is that torsion provides a geometric control of the orbital subband bottoms, while the Zeeman term separates the two spin-resolved branches. For low-bias Landauer transport, the relevant quantity is the band bottom $\Emin_{n,m,\sigma}(\tau)$ of each subband, which torsion lowers through $-\hbar^2\tau^2A_{nm}^2/2m^*$, so that channels open with increasing torsion. A spin-selective torsion interval appears when the Fermi level lies between the two spin-resolved band bottoms; its width, Eq.~\eqref{eq:spin_window}, is linear in $\Delta_Z$ in the small-splitting limit. The fixed-momentum energy $\varepsilon^{\mathrm{ch}}_{n,m,\sigma}(\tau;\kF)$ instead governs the vertical intersubband optical transition, and the two conventions are kept distinct throughout.

Finite-size and surface-boundary effects were incorporated by treating the device as a finite open scattering region of length $L$ and radius $R$. Longitudinal Fabry--Perot interference modulates the transmission, radial boundary conditions shift the absolute channel energies, and surface roughness introduces linewidth broadening. These effects do not remove the torsion-control mechanism provided the effective broadening satisfies $\Gamma_{\mathrm{eff}}\ll\Delta_Z$. This condition was quantified through polarization and conductance-contrast maps, which show the gradual loss of spin selectivity as $\Gamma_{\mathrm{edge}}$ approaches the Zeeman scale.

The same threshold framework also accounts for the optical and thermoelectric responses. Transverse intersubband absorption produces a THz resonance whose leading fixed-momentum slope is set by $\partial\omega/\partial\tau\simeq\hbar\kF/m^*$ for the dominant unit-branch transition, while finite-radius boundaries shift and broaden the absorption ridge. The spin thermopower changes sign as the torsion-shifted band bottoms move through the thermal window, providing a mechanical route to invert the thermally generated spin response. In addition, the shot-noise and Fano-factor response provide an independent transport fingerprint of the same torsion-driven channel openings through paired Zeeman-split peaks. Overall, the analysis identifies a compact organizing principle for torsion-controlled nanowires: geometric torsion lowers the transport band bottoms, Zeeman splitting resolves spin branches, fixed-momentum optical transitions retain a linear torsional slope, and surface-induced broadening sets the robustness scale.

\section*{Acknowledgments}
This work was supported by CAPES, CNPq, and FAPEMA. E.O.S. acknowledges support from grants CNPq/306308/2022-3, FAPEMA/UNIVERSAL-06395/22, and CAPES/Finance Code 001.

\end{document}